\documentclass[aps,pra,twocolumn]{revtex4-1}
\usepackage{amsfonts}
\usepackage{amsmath}
\usepackage{mathrsfs}
\usepackage{amssymb}
\usepackage{graphicx}
\usepackage{bm}
\usepackage{color}
\usepackage[colorlinks=true,linkcolor=blue,anchorcolor=blue,citecolor=blue,urlcolor=blue]{hyperref}

\begin{document}

\title{Spin-orbit-coupling-assisted roton softening and superstripes in a Rydberg-dressed Bose-Einstein Condensate}

\author{Hao Lyu}
\affiliation{International Center of Quantum Artificial Intelligence for Science and Technology (QuArtist) and Department of Physics, Shanghai University, Shanghai 200444, China}

\author{Yongping Zhang}
\email{yongping11@t.shu.edu.cn}
\affiliation{International Center of Quantum Artificial Intelligence for Science and Technology (QuArtist) and Department of Physics, Shanghai University, Shanghai 200444, China}

\begin{abstract}

Rotons can exist in ultracold atomic gases either with long-range interactions or with spin-orbit-coupled  dispersions. We find that two different kinds of rotons coexist in a joint system combining long-range interactions and spin-orbit coupling. One roton originates from spin-orbit coupling and two others come from long-range interactions. Their softening can be controlled separately. The interesting new phenomenon which we find is that spin-orbit-coupled roton can push down the energy of one long-range-interaction roton. The spin-orbit coupling accelerates the softening of this roton. The post phase of spin-orbit-coupling-assisted roton softening and instability is identified as a superstripe.

\end{abstract}

\maketitle

\section{Introduction}

In quantum many-body systems, roton is a particular kind of collective excitations and features a parabolic-like dispersion relation at a finite momentum. Supersolid~\cite{Penrose1956,Gross1957,Chester1970,Legget1970, Pomeau} is a unique phase of ground states and is characterized by the coexistence of crystalline densities due to spontaneous breaking of continuous translational symmetry and superfluidity associated with gauge symmetry breaking. These two different phenomena are connected by the so-called roton softening and instability. Varying relevant parameters gives rise to softening of the roton energy, and vanishing of the roton gap predicts that the homogeneous state hosting roton excitations is energetically unstable. Such a roton instability demands that the ground state possibly is a supersolid due to the simultaneous occupation of the original momentum and roton momentum where roton instability happens. Therefore, the roton softening and instability provides an accessible route to searching for supersolid phases~\cite{Ancilotto,Petter}.

Recently, there is a big advance for the study of these twin phenomena in ultracold atomic gases, produced by their experimental implementations.  In ultracold atoms, there are two different ``arenas''  to accommodate rotons and supersolids. One is with long-range interactions and the other is with special single-particle dispersion relations featuring multiple energy minima. Long-range interactions can form self-attractions around a certain momentum in momentum domain which generate a local minimum with infinite density of state, i.e., a roton~\cite{Santos,ODell,Henkel2010,Kora}.
Dominant long-range interactions can be experimentally realized with the assistance of optical cavities. Roton softening~\cite{Mottl} and supersolid phases~\cite{Esslinger2017} have been observed in pioneering optical cavity experiments. Very recently, experimental achievements of Bose-Einstein condensation in strongly magnetic lanthanide atoms provide another promising perspective to measure the dipolar roton softening~\cite{Petter} and identify supersolids~\cite{Ferlaino2018NaturePhys,Pfau2019PRX,Ferlaino2019PRX,Modugno2019PRL}, with the assistance of dominant long-range dipolar interactions. Furthermore, dipolar supersolids have been characterized by observations of their collective excitations from different aspects~\cite{Esslinger2018,Ferlaino2019PRL,Modugno2019Nature,Pfau2019Nature,Pfau2019PRL}.  Rydberg-dressed Bose-Einstein condensates (BECs) offer another platform to realize long-range interactions~\cite{Saffman2010,Bloch2015,Firstenberg2016,Browaeys2020}. They can be achieved by off-resonantly coupling BEC atoms to a Rydberg state~\cite{Henkel2010,Saffman2010}. Unlike anisotropic dipolar interactions, nonlocal Rydberg-dressed interactions can be isotropic. Supersolid phases have been identified theoretically in Rydberg-dressed BECs~\cite{Henkel2010,Henkel2012}. 

The other mechanism to introduce rotons and supersolids relates to single-particle dispersion relations. If single-particle dispersions possess multiple energy minima, the homogeneous ground state chooses to occupy one of them, another unoccupied minima are modified by repulsive atomic interactions to generate roton-like structures in collective excitations, no matter the interactions are short-range or long-range~\cite{Martone,Zheng2013,Khamehchi2014}. 
There already exists two experimental approaches to prepare such special single-particle dispersions. One utilizes periodically shaking optical lattices~\cite{Ha} and the other makes use of spin-orbit coupling induced by Raman coupling between lasers and atoms~\cite{Spielman2011,Spielman2013,Goldman,Zhai2015,Zhang2016}. Roton softening has been examined in both shaking lattices and spin-orbit coupled gases~\cite{Khamehchi2014,Pan2015,Ha}. Spin-orbit-coupled supersolids have also been observed experimentally in Ref.~\cite{Ketterle2017}. Because of their one-dimensional nature, they are called superstripes~\cite{Li2013}.

The combination of these two different arenas together endows rotons and supersolids with more interesting properties.  Supersolids and superstripes have been investigated in spin-orbit-coupled systems with long-range interactions in different spatial dimensions~\cite{YiSu,Clark,Demler,Lyu,Han2018}. Much attention has focused on supersolids. However, whether the rotons generated by the two different mechanisms can coexist and how the rotons in the combined arenas soften are not known yet.

In this paper, we address these questions by studying a spin-orbit-coupled BEC with long-range Rydberg-dressed interactions. In atomic BECs,  the pseudospin states of spin-orbit coupling could be hyperfine ground states.  The coupling of  hyperfine states by Raman lasers requires particular atoms~\cite{Spielman2009}. Currently, all spin-orbit-coupled BEC experiments with hyperfine states are implemented in $^{87}$Rb BECs~\cite{Spielman2011,Pan2015}. While, the possible realization of Rydberg-dressed interactions does not need to choose specific atoms~\cite{Saffman2010}, which provides a hope to implement both in$^{87}$Rb atoms. Until now, it is still a challenge to realize Rydberg-dressed BECs~\cite{Balewski2014}. However, Rydberg-dressed BECs attract lots of current theoretical interest as an outstanding platform to explore phenomena of long-range interactions~\cite{Seydi,Zhou2020}.

We find that there exists two different kinds of rotons separately originating from single-particle spin-orbit-coupled dispersion and long-range interactions. The softening of these rotons can be adjusted independently. This tunability introduces the system into a new stage where the spin-orbit-coupled roton can assist the softening of  long-range interaction-rotons. The consequence of such spin-orbit-coupling-assisted roton softening and instability is superstripe ground states. We characterize the superstripes by their collective excitation spectrum.  Our study is organized as follows. In Sec.~\ref{Theory}, we present our theoretical frame for the analysis of roton and superstripe. In Sec.~\ref{Softening}, we uncover that two kinds of rotons with different origination mechanism can coexist, and spin-orbit-coupling-assisted roton softening will be demonstrated. The post-roton-softening phase, i.e., superstripes, will be discussed in Sec.~\ref{Superstripe}, and the conclusion follows in Sec.~\ref{Conclusion}.

\section{Mean-field theory}
\label{Theory}

We start from a three dimensional atomic BEC and assume that harmonic traps along the transverse direction are strong enough so that the transverse motion of atoms is completely confined into the ground state of harmonic  traps. After integrating over the transverse motion, we get a quasi-one-dimension (1D) spin-orbit-coupled BEC with Rydberg-dressed interactions. The mean-field energy functional of the system is,
\begin{align}
\label{eq:energy}
E&=\int dx\psi^\dagger(x)H_{\mathrm{soc}} \psi(x)\notag\\
&\phantom{={}}+\frac{1}{2}\sum_{i,j=1,2}\int dx g_{ij}|\psi_i(x)|^2|\psi_j(x)|^2 \\
&\phantom{={}}+\frac{1}{2}\sum_{i,j=1,2}\int dxdx^\prime V_{ij}(x-x^\prime)|\psi_i(x)|^2|\psi_j(x^\prime)|^2.
\notag
\end{align}
Here $\psi=(\psi_1, \psi_2)^T$ is the two-component wave function. $g_{ij}$ ($i,j=1,2$) characterize contact interactions between intra- and inter-species, which are proportional to $s$-wave scattering lengths and the atom number.
$H_{\mathrm{soc}}$ is the single-particle spin-orbit-coupled Hamiltonian,
\begin{equation}
H_{\mathrm{soc}}=-\frac{1}{2}\frac{\partial^2}{\partial x^2}-i\gamma\frac{\partial}{\partial x}\sigma_z+\frac{\Omega}{2}\sigma_x.
\end{equation} 
In experiments, the spin-orbit coupling term $-i\gamma \partial/\partial x\sigma_z$ can be artificially introduced into atoms by momentum exchanges between Raman lasers and atoms~\cite{Khamehchi2014,Spielman2011,Pan2015}.
The spin-orbit coupling strength $\gamma=\hbar k_\mathrm{Ram}/m$ relates to the wavelength $\lambda_\mathrm{Ram}$ of Raman lasers with $k_\mathrm{Ram}=2\pi/\lambda_\mathrm{Ram} $, where $m$ is the atom mass. $\Omega$ is the Raman coupling and is proportional to the laser intensity, so that it can be easily tuned in experiments.  In all our dimensionless equations, we choose the unit of momentum, length, and energy as $k_\mathrm{Ram}$, $1/k_\mathrm{Ram}$ and $\hbar^2k_\mathrm{Ram}^2/m$, respectively. Therefore, under these units, the dimensionless parameter $\gamma=1$. However, the value of $\gamma$ can be varied by the periodic modulation of $\Omega$ according to the  proposal in Ref.~\cite{Zhang2013} and the experimental realization in~\cite{Jimenez}. The effective Rydberg-dressing potentials between intra- and inter-components are~\cite{Henkel2010,Saffman2010,Hsueh2013},
\begin{align}
V_{ij}(x)=\frac{\tilde{C}^{ij}_6}{x^6+R^6_c},
\end{align}
with $\tilde{C}^{ij}_6$ being the interaction strengths of Rydberg-dressing and $R_c$ being the blockade radius.

By minimizing the free energy $F=E-\mu N$ with $\mu$ being the chemical potential and $N$ being the total atom number, we get stationary Gross-Pitaevskii (GP) equations~\cite{Pomeau},
\begin{align}
\mu\psi= \left( H_{\mathrm{soc}}+H_{s}[\psi]+H_{\mathrm{Ryd}}[\psi]  \right) \psi.
\label{eq:GPE}
\end{align}
Here $H_{s}$ denotes the contact interactions,
\begin{align}
H_s[\psi]=\left(
\begin{matrix}
g_{11}|\psi_1|^2+g_{12}|\psi_2|^2 &  0\\
0 & g_{12}|\psi_1|^2+g_{22}|\psi_2|^2
\end{matrix}
\right). \notag
\end{align}
$H_\mathrm{Ryd}$ represents the long-range Rydberg-dressed interactions,
\begin{align}
&H_{\mathrm{Ryd}}[\psi]=\notag\\
&\sum_{i=1,2}\int dx^\prime
\begin{pmatrix}
V_{1i}(x^\prime-x)|\psi_i(x^\prime)|^2 &  0\\
0 & V_{2i}(x^\prime-x)|\psi_i(x^\prime)|^2
\end{pmatrix}. \notag
\end{align}
%
Once we know the ground-state wave function $\psi$ and corresponding $\mu$, we can study their collective excitations by adding perturbations into the ground state. Therefore total wave functions are,
\begin{align}
\label{eq:bogoliubov}
\Psi_{1,2}=e^{-i\mu t}\left[\psi_{1,2}(x)+u_{1,2}(x)e^{-i\omega t} +v^\ast_{1,2}(x)e^{i\omega t} \right],
\end{align}
where $\omega$ is the excitation energy, and $u_{1,2}(x), v_{1,2}(x)$ are  perturbation amplitudes, satisfying the normalization condition, $
\sum_{l=1,2}\int dx \left(|u_l(x)|^2-|v_l(x)|^2\right)=1$. After substituting $\Psi_{1,2}$ into the time-dependent version of Eq.~(\ref{eq:GPE}) and keeping linear terms of $u$ and $v$, we get Bogoliubov-de Gennes (BdG) equations~\cite{Dalfovo1999},
\begin{align}
\label{eq:BdG}
\mathcal{L}_1\phi+\mathcal{L}_2[\phi]=\omega \phi,
\end{align}
with $\phi=(u_1(x),u_2(x),v_1(x),v_2(x))^T$.
The matrix $\mathcal{L}_1$ is
\begin{align}
\mathcal{L}_1=
\left(
\begin{array}{cc}
H_\mathrm{soc}+A-\mu & B \\
B^\ast & -H_\mathrm{soc}-A^\ast-\mu
\end{array}
\right),
\end{align}
with
\begin{align}
A=H_s[\psi]+H_\mathrm{Ryd}[\psi]+
\left(
\begin{array}{cc}
g_{11}|\psi_1|^2 & g_{12}\psi^\ast_1\psi_2 \\
g_{12}\psi_1\psi^\ast_2 & g_{22}|\psi_2|^2
\end{array}
\right), \notag
\end{align}
and
\begin{align}
B=
\left(
\begin{array}{cc}
g_{11}\psi^2_1 & g_{12}\psi_1\psi_2 \\
g_{12}\psi_1\psi_2 & g_{22}\psi^2_2
\end{array}
\right). \nonumber
\end{align}
The matrix $\mathcal{L}_2[\phi]$ is given by
\begin{align}
\mathcal{L}_2[\phi]	
=\int dx^\prime
\left(
\begin{matrix}
M_1(x,x^\prime) & M_2(x,x^\prime) \\
M^\ast_2(x,x^\prime) & M^\ast_1(x,x^\prime)
\end{matrix}
\right)\phi(x^\prime),
\end{align}
with
\begin{align}
&M_1(x^\prime,x)=\nonumber\\
&\left(
\begin{matrix}
V_{11}(x^\prime-x)\psi^\ast_1(x^\prime)\psi_1(x) & V_{12}(x^\prime-x)\psi^\ast_2(x^\prime)\psi_1(x) \\
V_{12}(x^\prime-x)\psi^\ast_1(x^\prime)\psi_2(x) & V_{22}(x^\prime-x)\psi^\ast_2(x^\prime)\psi_2(x)
\end{matrix}
\right), \notag
\end{align}
\begin{align}
&M_2(x^\prime,x)=\notag\\
&\left(
\begin{matrix}
V_{11}(x^\prime-x)\psi_1(x^\prime)\psi_1(x) & V_{12}(x^\prime-x)\psi_2(x^\prime)\psi_1(x) \\
V_{12}(x^\prime-x)\psi_1(x^\prime)\psi_2(x) & V_{22}(x^\prime-x)\psi_2(x^\prime)\psi_2(x)
\end{matrix}
\right).\notag
\end{align}

In the following, we solve GP equations~(\ref{eq:GPE}) to search for ground states and solve BdG equations~(\ref{eq:BdG}) to analyze the collective excitation spectrum of corresponding ground states. We set  $g_{11}=g_{22}=g_{12}=g$, since they are approximately the same in experiments~\cite{Spielman2011}. The two-component Rydberg-dressed BEC may be realized by coupling two ground hyperfine states to different Rydberg states~\cite{Han2018,Hsueh2013}.  
The strength of long-range interactions  $\tilde{C}^{ij}_6$ and the blockade radius $R_c$ depend on the two-photon detuning and Rabi frequency of excitation lasers, which are tunable in experiments \cite{Saffman2010}. In this work, we choose $V_{11}(x)=V_{22}(x)=V_{12}(x)=V(x)$ for simplicity, so that $\tilde{C}^{ij}_6= \tilde{C}_6$. 
For further convenience, we transform $V(x)$ into the momentum space, i.e., $\tilde{V}(k)=\int dx V(x) \exp(ikx)=\tilde{V}_0f(k)$, with $\tilde{V}_0=2\pi\tilde{C}_6/3R^5_c$ and 
$f(k)=\frac{1}{2}e^{-|k|Rc/2}\left[e^{-|k|R_c/2}+\cos\left(\sqrt{3}|k|R_c/2 \right) \right.
\phantom{={}}+\left.\sqrt{3}\sin\left(\sqrt{3}|k|R_c/2 \right)\right]
$. We use parameters $\tilde{V}_0$ and $R_c$ to characterize the Rydberg-dressed interactions. The validity of theoretical mean-field frame in our quasi-1D system is examined by the calculation of quantum depletion demonstrated the appendix.

\section{Spin-orbit-coupling-assisted roton softening}
\label{Softening}

\begin{figure}[t]
\centering
\includegraphics[width=2.3in]{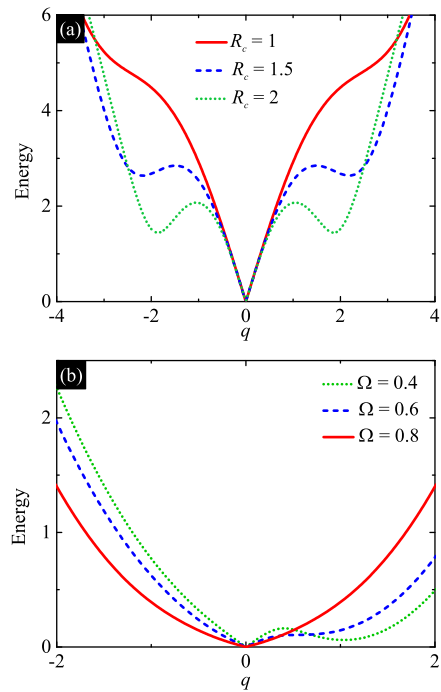}
\caption{(a) The density excitation of a Rydberg-dressed BEC for various $R_c$ with a fixed $\tilde{V}_0=10$. (b) The lower branch of the collective excitation of the plane-wave ground state in a spin-orbit coupled BEC without long-range interactions. The spin-orbit coupling strength is set as $\gamma=0.6$. In both (a) and (b), $s$-wave interaction strength is fixed as $g=0.5$.}
\label{fig1}
\end{figure}

Without the spin-orbit coupling ($\gamma=0$, $\Omega=0$), it is known that the collective excitation spectrum $\omega(q)$ of a BEC with long-range interactions can be analytically calculated from the BdG equations~\cite{Wilson}. In collective excitations, there are two branches~\cite{Abad,Liang2018}; one is spin-density excitations $\omega=q^2/2$, and the other is density excitations,
\begin{align}
\omega=\frac{1}{2}\sqrt{q^4+4q^2\left[g+\tilde{V}(q)\right]}.
\label{eq:rydberg}
\end{align}
The density branch is consistent with the collective excitation spectrum of a single component with long-range interactions~\cite{Henkel2010}. In Fig.~\ref{fig1}(a),  the density branch as a function of $R_c$ is shown for a fixed $\tilde{V}_0$. We check that $\tilde{V}(q)$ becomes attractive around $|q|\sim4.3/R_c$.
It predicts~\cite{Ancilotto} that the roton momentum $q_\mathrm{rot}$ where minimum of roton energy happens can be characterized by the blockade radius, which is consistent with the finding in Ref.~\cite{Henkel2010}.  One of the interesting features of Rydberg-dressed rotons is that rotons appear in a pair and symmetrically distribute at $q_\mathrm{rot}\sim \pm 4.3/R_c $. Increasing $R_c$ leads to the shrinking of roton minimum and meanwhile the reduction of roton gap. Thus, roton softening can be introduced by increasing $R_c$ [see Fig.~\ref{fig1}(a)]. Once the roton gap is closed, roton instability happens and further increasing $R_c$ leads to the roton energy becoming complex-valued. 

As mentioned above, we unravel the mechanism of the existence of roton softening in a Rydberg-dressed BEC. In the following, we show that roton softening can exist in a spin-orbit-coupled BEC. The dispersion relation of the single-particle spin-orbit-coupled Hamiltonian $H_{\mathrm{soc}}$ possesses two bands and the lowest band has two degenerate energy minima. These two minima give rise to the roton spectrum. In order to show this, we assume that the ground state is a plane-wave, so that the wave functions are $\psi(x)=e^{ikx}(\varphi_1,\varphi_2)^T$~\cite{Martone,Li2012} with $k$ being the ground-state quasimomentum. $\varphi_{1,2}$ are independent of spatial coordinates and satisfy the renormalization condition $|\varphi_1|^2+|\varphi_2|^2=1$. 
With this ansatz, the energy functional becomes
\begin{align}
E&=\frac{1}{2}k^2+\gamma k(|\varphi_1|^2-|\varphi_2|^2)+\frac{\Omega}{2}(\varphi^\ast_1 \varphi_2+ \varphi_1\varphi^\ast_2) \nonumber\\
&\phantom{={}}+\frac{1}{2}(g+\tilde{V}_0). \notag
\end{align}
From the above expression, it is clear that long-range and contact interactions contribute only an overall shift of the energy since all their coefficients are the same.
This leads to the fact that the properties of the ground state should be similar to that of the single-particle case. 
For $\Omega< 2\gamma^2$, minimization of the energy functional results in two energy minima laying at quasimomentum $k_\pm=\pm\gamma\sqrt{1-\Omega^2/4\gamma^2}$, and the plane-wave ground state spontaneously chooses one of them to occupy.
By using the results of the minimization with the calculated chemical potential and assuming
$u_{1,2}(x)=u_{1,2}e^{i(k+q)x}$ and $v_{1,2}(x)=v_{1,2}e^{-i(k-q)x}$ \cite{Martone} with $q$ being the perturbation quasimomentum, we calculate the BdG equations to obtain the collective excitation spectrum $\omega(q)$ of the plane-wave ground state. Figure~\ref{fig1}(b) demonstrates roton softening by varying the Raman coupling $\Omega$ without long-range interactions ($\tilde{V}_0=0$). The spin-orbit-coupled roton only exists in one side of $q$. This is because the plane-wave ground state that we use to plot Fig.~\ref{fig1}(b) spontaneously occupies $k_-$. 
Since the contact interactions $g$ contributes positive energy to the system, they displace single-particle dispersion upwards and meanwhile shape the dispersion around $k_-$ (corresponding to $q=0$ in the perturbation momentum frame) into a linear form. A phonon mode can be formed in this mechanism.
Therefore, the other unoccupied single-particle minimum at $k_+$ appears as a roton-like structure in collective excitations. In the perturbation quasimomentum frame, a roton sits at $q_\mathrm{rot}\approx k_+-k_-=2\gamma\sqrt{1-\Omega^2/4\gamma^2}$.  Decrease of $\Omega$ causes roton softening, as well as the increase of roton momentum [see Fig.~\ref{fig1}(b)]. 
Further decrease of $\Omega$ can not result in roton instability. This is because there is no phase transition from the plane-wave ground state to superstripes,
with all equal interaction coefficients~\cite{Li2012}. If we make the interaction coefficients unequal, roton instability occurs after the softening and it features negative roton energy~\cite{Khamehchi2014}.

\begin{figure}[t]
\centering
\includegraphics[width=2.3in]{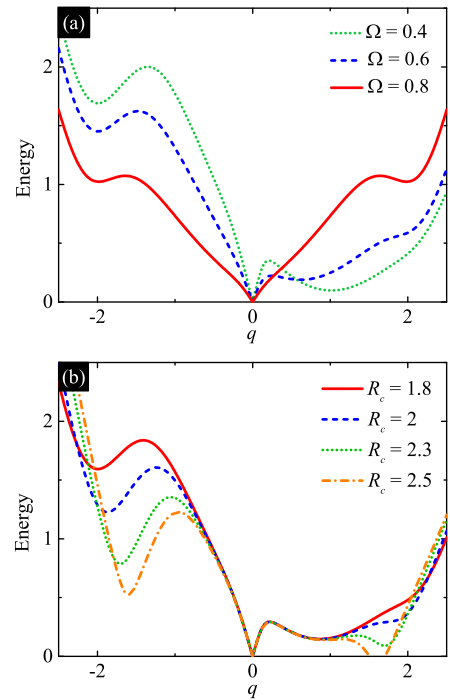}
\caption{Collective excitation spectrum of the Rydberg-dressed BEC with spin-orbit coupling, in which we choose $\gamma=0.6$, $g=0.5$, and $\tilde{V}_0=10$. (a) Evolution of the spectrum for various $\Omega$ with a fixed $R_c=1.8$. (b) Evolution of the spectrum for various $R_c$ with a fixed $\Omega=0.5$. For the line with $R_c=2.5$, the right-hand Rydberg-dressed roton become unstable with complex-valued energy which is not shown in the plot.}
\label{fig2}
\end{figure}

\begin{figure*}[t]
\centering
\includegraphics[width=4.9in]{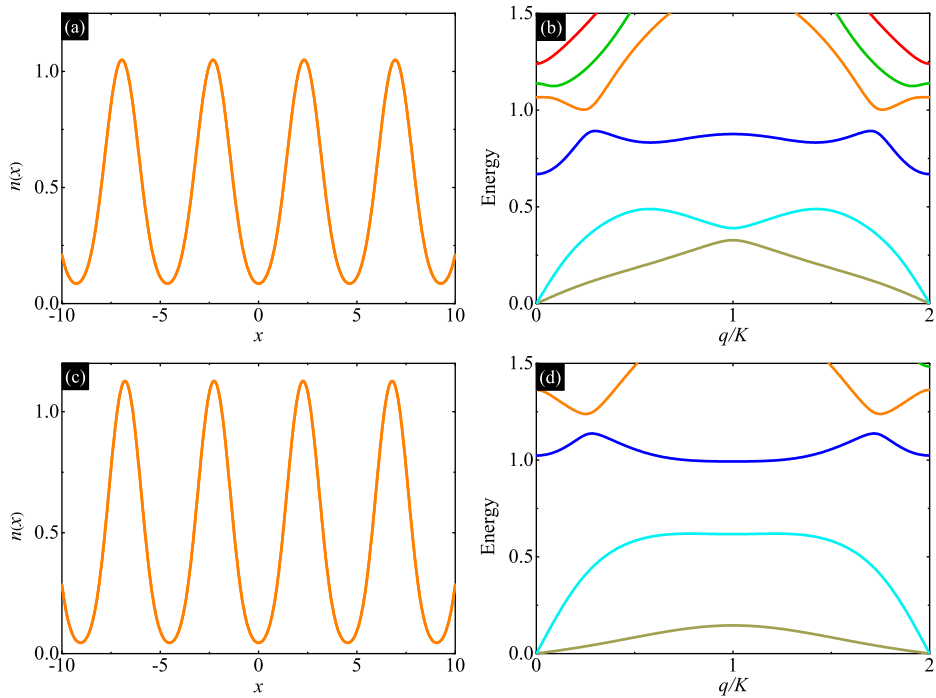}
\caption{Density profiles (a,c) and excitation spectrums (b,d) of the superstripe phase. We choose $\Omega=0.4$ in (a,b) and 0.8 in (c,d). The other parameters are $\gamma=0.6$, $g=0.5$, $\tilde{V}_0=10$, and $R_c=3$. }
\label{fig3}
\end{figure*}

We conclude that the momentum of Rydberg-dressed roton is governed by the blockade radius $R_c$, while the spin-orbit-coupled-roton's momentum relates to spin-orbit coupling parameters $\gamma$ and $\Omega$.  They have completely different originations, which intuitively makes their coexistence possible. We perform numerical calculations to confirm this expectation.  The collective excitation of a BEC in the presence of both spin-orbit coupling and long-range Rydberg-dressed interactions are demonstrated in Fig.~\ref{fig2}(a). For $R_c=1.8$, we see that there are only Rydberg-dressed rotons sitting at $q_\mathrm{rot}\sim \pm 4.3/R_c \sim \pm 2.2$ symmetrically when $\Omega$ is large [see the line with $\Omega=0.8$ in Fig.~\ref{fig2}(a)]. This is expected since for a larger $\Omega$ there is no spin-orbit-coupled roton even without long-range interactions [see the line with $\Omega=0.8$ in Fig.~\ref{fig1}(b)]. Decreasing $\Omega$ leads to the appearance of the spin-orbit-coupled roton sitting approximately at 
$2\gamma\sqrt{1-\Omega^2/4\gamma^2}$. The newborn roton pushes down the spectrum at right-hand side, and pushes up the left-hand side spectrum. However, it does not affect the location of Rydberg-dressed rotons [see the lines with $\Omega=0.6$ and $0.4$ in Fig.~\ref{fig2}(a)]. From the line with $\Omega=0.4$ in Fig.~\ref{fig2}(a), where the spin-orbit-coupled roton dominates, the Rydberg-dressed roton at right side fades away while the left side roton survives. 

Consequently, the two different kinds of rotons can exist independently. Meanwhile,  the spin-orbit-coupled roton dramatically pushes spectrum on its side down, which provides a new phenomenon for the softening of the Rydberg-dressed roton. In Fig.~\ref{fig2}(b) we show the Rydberg-dressed roton softening by varying $R_c$. We start from the coexistence case with $R_c=1.8$ (the red line). Due to the spin-orbit-coupled roton, the spectrum is completely asymmetric with respect to $q=0$. The right-hand side Rydberg-dressed roton is obviously lower than the left-hand one. Surprisingly, the increase of $R_c$ quickly brings the right-hand roton into the unstable situation, while the left-hand one softens but still has a large roton gap [see the line with $R_c=2.5$ in Fig.~\ref{fig2}(b)]. Very interestingly, the change of $R_c$ does not affect the spin-orbit-coupled roton. 

Therefore, we can have parameter regimes in which only one Rydberg-dressed roton becomes unstable. This is fully distinguishable with the case without spin-orbit coupling where paired Rydberg-dressed rotons soften and are unstable in the same time. The reason of one unstable roton is that it is the spin-orbit-coupled roton that helps to lower its energy to prepare for the occurring of instability. Thus, the spin-orbit coupling plays a role of an assistance for one Rydberg-dressed roton softening and instability.  It can enhance the roton softening and accelerate instability to happen. We emphasize that for our parameters the spin-orbit-coupled roton can not become unstable and the instability always happens in the Rydberg-dressed rotons.

\begin{figure*}[t]
	\centering
	\includegraphics[width=6.1in]{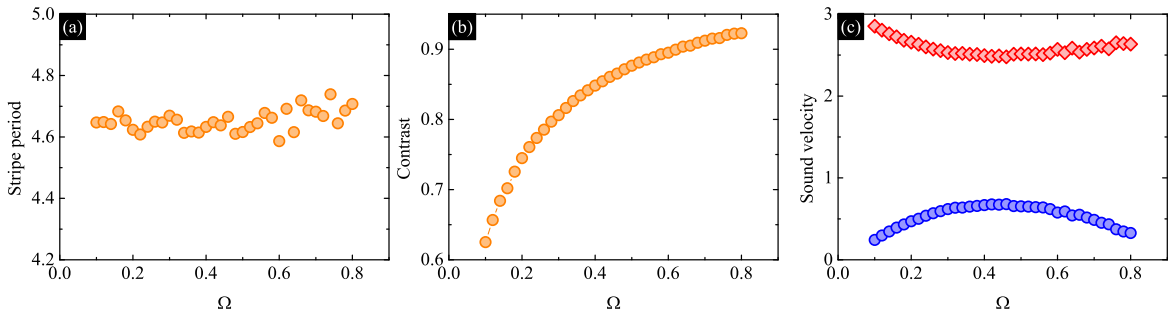}
	\caption{The period (a) and contrast (b) of the superstripes as a function of the Raman coupling strength $\Omega$. (c) Sound velocity of the upper (circles) and lower (squares) phonon branches as a function of the Raman coupling strength $\Omega$. The parameters are $\gamma=0.6$, $g=0.5$, $R_c=3$, and $\tilde{V}_0=10$.}
	\label{fig4}
\end{figure*}

\section{Post-roton-instability phase: superstripe}
\label{Superstripe}

Now we are in the position to ask: what are properties of post-roton-instability phase induced by such a new mechanism of the spin-orbit-coupling-assisted roton softening.  To address this question, we numerically find the existence of superstripe ground state. Considering the periodic density distributions of the superstripe, we assume that the wave functions are periodic and can be expressed in a plane-wave basis~\cite{Li2013,Hu2018},
\begin{align}
\psi(x)=
\sum^{L}_{n=-L}
\left( 
\begin{matrix}
\psi^{(n)}_1\\
\psi^{(n)}_2
\end{matrix}
\right) e^{i nKx },
\label{eq:stripe}
\end{align}
where $\psi^{(n)}_1$ and $\psi^{(n)}_n$  satisfy the renormalization condition $\sum_n(|\psi^{(n)}_1|^2+|\psi^{(n)}_2|^2)=1$, and $L$ is the cutoff of the plane-wave modes. This ansatz indicates that the period of wave functions is relevant to $K$, whose value shall be determined by minimizing corresponding energy functional obtained by substituting Eq.~(\ref{eq:stripe}) into Eq.~(\ref{eq:energy}). Minimization procedure demonstrates that only the coefficients of odd-numbered plane-wave modes (i.e., $e^{\pm iKx}, e^{\pm i3Kx}$, etc.) are nonzero. Therefore, the period of wave functions is numerically exact $2\pi/K$ and  the density period  $\pi/K$. These results are similar to that in spin-orbit-coupled BECs without Rydberg-dressed interactions~\cite{Li2013,Hu2018}. Two typical density distributions are shown in Fig.~\ref{fig3}(a,c) for $\Omega=0.4$ and $0.8$ respectively. We see that $\Omega$ has little impact on density periodicity. The density period $\pi/K$ is $\sim 4.6$, which is consistent with the predicted value from the roton instability. For $R_c=3$ which is used in Fig.~\ref{fig3}, we check the
roton instability centered around $q_\mathrm{rot}\sim 1.36$, which predicts that the period of post-roton-instability phase should be $2\pi/q_\mathrm{rot} \sim 4.6$. The density period for various $\Omega$ is demonstrated in Fig.~\ref{fig4}(a). It confirms that density period is insensitive to $\Omega$.

 
 We further characterize the superstripe ground states by their collective excitation spectrum. It is calculated by substituting the superstripe wave functions into BdG equations~(\ref{eq:BdG}) and using the following perturbation amplitudes,
 \begin{align}
 u_{1,2}(x)&=e^{iqx}\sum^{L}_{n=-L}u^{(n)}_{1,2}e^{i(2n+1)Kx},\notag\\
 v_{1,2}(x)&=e^{iqx}\sum^{L}_{n=-L}v^{(n)}_{1,2}e^{i(2n+1)Kx}.
 \end{align}
The reason for the choice of above amplitudes is that the BdG equations are periodic with the period being $\pi/K$, as a result of the periodic density of the superstripes. Therefore, the amplitudes should be Bloch states with $q$ being perturbation quasimomentum and we expand their periodic parts by a plane-wave basis. The results obtained by diagonalizing the BdG equations are shown in Fig.~\ref{fig3}(b,d), corresponding to the superstripes plotted in Fig.~\ref{fig3}(a,c) respectively. There is a slight difference between the two superstripes in Fig.~\ref{fig3}(a,c). But the collective excitation spectrum are obviously different, which features a Bloch band-gap structure. The gap sizes in Fig.~\ref{fig3}(d) is clearly larger than these in Fig.~\ref{fig3}(b). This is because the density of superstripe in Fig.~\ref{fig3}(c) have a higher amplitude than that in Fig.~\ref{fig3}(a). A high density will open a large gap in collective excitations. We label the density amplitude by the contrast of superstripe, which is defined as
\begin{align}
C=\frac{n_{\mathrm{max}}-n_{\mathrm{min}}}{n_{\mathrm{max}}+n_{\mathrm{min}}},
\end{align}
with $n_\mathrm{max}$ and $n_\mathrm{min}$ being the maximum and minimum of density respectively. The contrast as a function of $\Omega$ is plotted in Fig.~\ref{fig4}(b). It increases with the increase of the $\Omega$. Thus, we expect that a larger Raman coupling gives rise to larger gap sizes in collective excitations. 
 

In the long wavelength regime, the lowest two bands of collective excitation spectrum are phonon modes with different sound velocities. The existence of these two gapless Goldstone modes~\cite{Li2013,Saccani} is guaranteed by two spontaneous symmetry breaking of superstripes, one of which is continuous translational symmetry breaking and the other is gauge symmetry breaking. Sound velocities relate to slopes of phonon modes are analyzed as a function of $\Omega$, and the result is shown in Fig.~\ref{fig4}(c). The velocity of the upper phonon branch decreases slowly with increasing $\Omega$ [see circles in Fig~\ref{fig4}(c)]. However, the velocity of the lower phonon branch slowly increases and then decreases as a function of $\Omega$ [see squares in Fig~\ref{fig4}(c)].  Sound velocity shows a jump around phase transition from plane-wave to superstripe phases.  Therefore, the phase transition is first order.

Finally, we emphasize that the superstripes studied above are ground states supported by spin-orbit-coupling-assisted softening and instability. Without spin-orbit coupling, the ground states of relevant parameters are density homogeneous with roton excitations.

\section{Conclusion}
\label{Conclusion}

Long-range Rydberg-dressed interactions and spin-orbit coupling can separately generate collective excitations with roton structures. In a system with both Rydberg-dressed interactions and spin-orbit coupling, two different kinds of rotons from different originations can coexist. The location and softening of these rotons are adjustable independently. The interplay of them leads to an interesting phenomenon that the spin-orbit-coupled roton can assist Rydberg-dressed roton softening.  
The post-roton-instability phase is a superstripe, which is identified by analyzing their collective excitations. 
We conclude that spin-orbit coupling provides a possible means to accelerate roton softening.

\section*{Acknowledgement}

We thank Zhaoxin Liang and Zhu Chen for the useful help. This work is supported by the NSF of China (Grant Nos. 1174219 and 11974235), the Thousand Young Talents Program of China, and the Eastern Scholar and Shuguang (Program No. 17SG39) Program. H.~L. acknowledges the support by China Postdoctoral Science Foundation (Grant No. 2019M661457).

\appendix

\section{Quantum depletion}

\begin{figure}[t]
\centering
\includegraphics[width=2.3in]{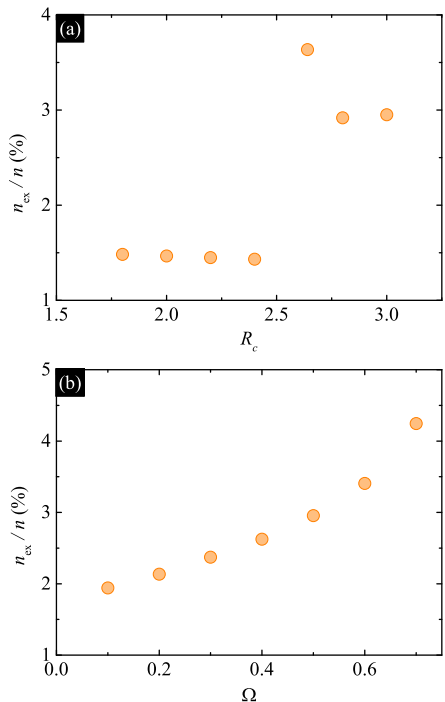}
\caption{Quantum depletion $n_\mathrm{ex}/n$ of the Rydberg-dressed BEC with spin-orbit coupling. 
(a) Quantum depletion as a function of $R_c$ with a fixed $\Omega=0.5$. (b) Quantum depletion as a function of $\Omega$ with a fixed $R_c=3$.
Other parameters are $\gamma=0.6$, $g=0.5$, $\Omega=0.5$, $\tilde{V}_0=10$, and $n=200k_\mathrm{Ram}$. A cutoff $q_c=0.01k_\mathrm{Ram}$ is introduced to prevent divergence.}
\label{fig5}
\end{figure}
In atomic BECs, quantum fluctuation induces a fraction of the condensate depleted even at zero temperature. The mean-field theory is valid when the number of depleted atoms is much smaller than the total atom number. Quantum depletion has been calculated in spin-orbit-coupled BECs~\cite{Hu2018,Ozawa,Cui,Chen,Liang}. In the following, we  show that quantum depletion of our system is small and the mean-field description is reasonable.

We start from three-dimensional GP equations and consider the atoms are strongly confined in the transverse direction. 
The transverse wave functions can be assumed to have a Gaussian shape. By integrating the transverse motion, we can obtain the quasi-1D GP equations. 
At zero temperature, the density of depleted atoms can be calculated as \cite{Hu2018}
\begin{align}
n_\mathrm{ex}=\sum_{j}\sum_{l=1,2}\sum_{q\neq0}\int dx\left|v^{(j)}_{q,l}(x)\right|^2.
\end{align}
Here, the superscript $j$ labels the $j$th Bogoliubov band.
The perturbation amplitude $v_{q,l}^j(x)$ depends on the quasimomentum $q$, which is given in the main text. Here, the transverse excitations are neglected due to the strong confinement. In Fig.~\ref{fig5}(a), we show that by tuning the Rydberg blockade radius, the quantum depletion $n_{\mathrm{ex}}/n$ jumps around the phase transition from plane-wave ($R_c<2.5$) to superstripe  ($R_c>2.5$) phases. $n$ is the BEC density, here we use a value $n=200k_\mathrm{Ram}$ that is typical in usual BEC experiments.  In the calculation, a cutoff of $q$ must be introduced to avoid divergence at very close to $q=0$~\cite{Gaul}, here the cutoff we used is $q_c=0.01k_\mathrm{Ram}$. For the superstripe phase, the quantum depletion increase with the increase of the Rabi frequency, as shown in Fig.~\ref{fig5}(b).
In our parameter regimes, $n_{\mathrm{ex}}/n$ is less than 5\%, which indicates the validity of our mean-field theory.

\end{document}